\documentclass{elsart}

\usepackage{amssymb}
\usepackage{amsmath}
\usepackage{booktabs}
\usepackage{verbatim}
\usepackage{pdflscape}
\usepackage{graphicx}
\usepackage{epsfig,amsmath,amssymb,graphicx,lscape,color}
\usepackage{multirow}
\usepackage{color}

\journal{Elsevier}
\begin{document}

\begin{frontmatter}

\title{The role of initial geometry in experimental models of wound closing}
\author{Wang Jin$^{1}$, Kai-Yin Lo$^{2}$, Shih--En Chou$^{2}$, }
\author{Scott W McCue$^{1}$}
\author{$^*$Matthew J Simpson$^{1}$}
\corauth[cor]{Corresponding author}
\address{$^1$ School of Mathematical Sciences, Queensland University of Technology (QUT) Brisbane, Queensland 4000, Australia.}
\address{$^2$ Department of Agricultural Chemistry, National Taiwan University\\
Taipei 10617, Taiwan.}
\ead{matthew.simpson@qut.edu.au, \textit{\textrm{Telephone}} +
617 31385241, \textit{\textrm{Fax}} + 617 3138 2310}

\begin{abstract}
Wound healing assays are commonly used to study how populations of cells, initialised on a two-dimensional surface, act to close an artificial wound space.  While real wounds have different shapes, standard wound healing assays often deal with just one simple wound shape, and it is unclear whether varying the wound shape might impact how we interpret results from these experiments. In this work, we describe a new kind of wound healing assay, called a \textit{sticker assay}, that allows us to examine the role of wound shape in a series of wound healing assays performed with fibroblast cells. In particular, we show how to use the sticker assay to examine wound healing with square, circular and triangular shaped wounds. We take a standard approach and report measurements of the size of the wound as a function of time.  This shows that the rate of wound closure depends on the initial wound shape.  This result is interesting because the only aspect of the assay that we change is the initial wound shape, and the reason for the different rate of wound closure is unclear.  To provide more insight into the experimental observations we describe our results quantitatively by calibrating a mathematical model, describing the relevant transport phenomena, to match our experimental data.    Overall, our results suggest that the rates of cell motility and cell proliferation from different initial wound shapes are approximately the same, implying that the differences we observe in the wound closure rate are consistent with a fairly typical mathematical model of wound healing. Our results imply that parameter estimates obtained from an experiment performed with one particular wound shape could be used to describe an experiment performed with a different shape.  This fundamental result is important because this assumption is often invoked, but never tested.
\end{abstract}

\end{frontmatter}

\newpage

\section{Introduction}
Two--dimensional \textit{in vitro} cell migration assays are routinely used to study wound healing and cancer cell spreading.  In these assays, both cell migration and cell proliferation play a key role at the phenotypic level (Keese et al., 2004; Singh et al., 2003). Generally there are two types of cell migration assays: (i) \textit{Proliferation assays} are initiated by placing cells as a uniform monolayer, at low density, onto a two-dimensional surface (Jones et al., 2001); and (ii) \textit{Wound  healing assays} are initiated by creating an artificial wound in a uniformly distributed monolayer of cells (Kramer et al., 2013).  In a proliferation assay, individual cells undergo both migration and proliferation events, which increases the cell density in the spatially uniform monolayer of cells (Jones et al., 2001). In contrast, wound healing assays involve individual cells moving into the initially vacant wound area.  The effects of cell migration and cell proliferation, combined, lead to the eventual closure of the wound space (Kramer et al., 2013).

Various types of wound healing assays are reported in the literature (Ariano et al., 2011; Ascione et al., 2017; Cai et al., 2007; Gough et al., 2011; Keese et al., 2004; Lee et al., 2010; Sheardown and Chent, 1996; Treloar et al., 2013).  Despite the fact that real wounds can take on arbitrary shapes and sizes, the most common wound shapes used in \textit{in vitro} wound healing assays are a simple long thin rectangular wound or a circular wound (Keese et al., 2004; Riahi et al., 2012; Tremel et al., 2009; van der Meer et al., 2010; Yarrow et al., 2004) and the question of whether the initial wound shape plays an important role is often overlooked (see Table \ref{WoundAssaySummary}).    We note that Madison and Gronwall (1992) compare the rate of healing for circular and square skin wounds on the dorsum of the metacarpus of horses, and they suggest that the rate of wound healing is not affected by the initial wound shape.  However, more recent quantitative analysis of several \textit{in vitro} models of wound healing suggest that the rate of wound healing can be extremely sensitive to the initial configuration of cells (Jin et al., 2016, Treloar et al., 2014).  Therefore, it is of interest to develop a new \textit{in vitro} assay which can be used to study wound healing for a variety of wound shapes, and to analyze results from the new experiments using a mathematical model to provide quantitative insight into the role of initial wound shape.

\begin{landscape}
\begin{table}
\caption{\bf{A summary of various types of wound healing assays.}}
\renewcommand{\arraystretch}{1.2}
\centering
\begin{tabular}{|l|l|l|l|}
\hline
    \textbf{Assay type} & \textbf{Wound shape} & \textbf{Advantages} & \textbf{Limitations}\\
    \hline
    Scratch assay & Long, thin rectangle & Straightforward to perform & Disturb surrounding cells (Jin et al., 2017) \\
    & (Jin et al., 2016) & (Liang et al., 2007) & Disrupt substrate (Liang et al., 2007)\\
    & & Relatively economic (Liang et al., 2007) & Not well--defined scratch areas \\
    & &  &  (Gough et al., 2011)\\
    \hline
    Barrier assay & Rectangle (Vedula et al., 2013)& Can vary the spreading & May require dry substrate \\
    & Circle (Treloar et al., 2013) & direction (Treloar et al., 2013) & (Ariano et al., 2011)\\
    \hline
    Electric impedance & Rectangle  & Monitor wound automatically  & Require special plates (Keese et al., 2004)\\
    assay & (Mamouni and Yang, 2011) &  (Keese et al., 2004) & \\
    & Circle (Keese et al., 2004) &  & \\
    \hline
    Stamp assay & Arbitrary (Lee et al., 2010) & Arbitrary wound shapes (Lee et al., 2010) & Require specific stamps (Lee et al., 2010)\\
    & & Presence of cell debris (Lee et al., 2010) &  \\
    \hline
    Sticker assay & Arbitrary & Arbitrary wound shapes & Requires laser scriber\\
    & & Straightforward to perform &  \\
    \hline
\end{tabular}
\label{WoundAssaySummary}
\end{table}
\end{landscape}

Experimental data from \textit{in vitro} wound healing assays are commonly presented by plotting the time evolution of wound area as the experiment proceeds and the wound closes (Bachstetter et al., 2016; Johnston et al., 2015; Ueck et al., 2017; Yarrow et al., 2004). While some studies report the time evolution of the exact wound area (Leu et al., 2012; Ueck et al., 2017; Yarrow et al., 2004), others studies report results in a non-dimensional format by reporting the wound area relative to the initial wound area (Ascione et al., 2017; Bachstetter et al., 2016; Johnston et al., 2015;  Katakowski et al., 2017; Walter et al., 2010). Although both types of measurements provide an indication of the speed at which wound healing takes place, reporting the data in terms of the relative wound area does not provide any information about the role of initial wound shape or initial wound size. Therefore, when comparing wound healing assays with different initial wound shapes, we believe it is important to report the data in terms of the wound area because this explicitly accounts for differences in the initial condition instead of simply reporting the area data relative to the initial wound area.

Many different types of mathematical and computational models have been used to mimic \textit{in vitro} collective cell migration assays. One approach is to use a discrete random walk model (Codling et al. 2008). In some random walk models, cell migration is represented by an unbiased, nearest-neighbour exclusion process, in which cell-to-cell crowding is modelled by hard core exclusion (Painter and Hillen, 2002; Simpson et al., 2010). Cell proliferation can be modelled by allowing individual agents in the simulation to divide to produce daughter agents (Simpson et al., 2010). Crowding effects can be incorporated into the proliferation mechanism by randomly choosing a nearest neighbour lattice site for the placement of the daughter agent, and only allowing the proliferation event to succeed if the target site is vacant (Simpson et al., 2010).  While discrete random walk models provide information relevant to individual cells within the population, it is also possible to describe the behaviour of the population of cells by considering the continuum-limit description of the random walk model, which in this case, gives rise to a two-dimensional reaction diffusion partial differential equation (PDE) that is equivalent to the two-dimensional Fisher-Kolmogorov model (Fisher, 1937; Kolmogorov et al. 1937).  This connection with the Fisher-Kolmogorov model is of interest because this model, and many other generalisations of this model, have been used previously to study collective cell migration problems, including wound healing type assays (e.g. Maini et al., 2004a, 2004b; Painter and Sherratt, 2003; Sengers et al., 2007; Sherratt and Murray, 1990; Swanson et al., 2003; Swanson, 2008).

In this study we develop and describe a novel experimental approach to investigate the role of initial wound shape.  We perform sticker assays using fibroblast cells and three different initial wound shapes: squares, circles, and equilateral triangles.  In addition to describing our new experimental protocol, and experimental results, we attempt to quantify the mechanisms that drive the wound healing process by calibrating the solution of a mathematical model to match the experimental data.  To estimate the rate of cell proliferation and the carrying capacity density, we examine a series of proliferation assays and apply the continuum-limit description of the random walk model to match the experimental data.  To estimate the cell diffusivity we use the discrete random walk model to mimic a series of sticker assays with different shaped wounds.   Comparing the snapshots from discrete simulations to the experimental images allows us to choose the cell diffusivity so that the discrete model matches the experimental images in terms of the position of the leading edge of the population of cells.  Overall, our results indicate that the parameters obtained from different wound shapes are approximately constant, suggesting that the initial wound geometry has no identifiable impact on the key mechanisms of cell motility and cell proliferation.  This is an important outcome because it is common to perform a wound healing assay with one particular geometry and to simply assume that the results might apply to another geometry, and this standard assumption is rarely considered or tested.  Furthermore, this result is important because our experimental data, alone, shows that the rate of wound closure depends on the initial shape of the wound.

Overall, while we find that different initial wound shapes lead to different rates of wound closure, our careful calibration of a mathematical model to the experimental data confirms that the differences in observed wound closure rates are entirely consistent with the underlying transport phenomena that drives wound healing.  In particular, we find that the rates of cell migration and cell proliferation are unaffected by the initial wound shape.

\section{Methods}
\subsection{Experimental methods}
The new experimental protocol for the sticker assay is shown schematically in Figure 1.  We perform the experiments with NIH 3T3 fibroblast cell line purchased from the Bioresource Collection and Research Center (BCRC), Taiwan. A complete medium composed of Dulbecco's Modified Eagle's medium (DMEM, Gibco, USA) and 10\% calf serum (CS, Invitrogen, USA) is used for cell culture. Cells are incubated in tissue culture polystyrene (TCPS) flasks (Corning, USA) in 5\% CO$_2$ at 37 $^{\circ}$C, and grown to approximately 90\% confluence before each passage.

The wound shapes are drawn in AutoCAD (Autodesk, USA) and then  loaded into a CO$_2$ laser scriber (ILS2, Laser Tools \& Technics Corp., Taiwan), to ablate desired wound shapes on a double-sided sticker (8018, 3M, USA). Three different types of wound shapes are designed in this experiment: square, circle, and equilateral triangle. The side length of each square and triangle sticker is 2~mm, and the diameter of each circle sticker is also 2~mm. The sticker is folded to form a handle, and is attached to the centre of a dish, with a diameter 35~mm. The dish is exposed to UV for 30 minutes for sterilisation. 3 $\times$ 10$^5$ cells are placed, as uniformly as possible, into the dish, and incubated overnight.  To initiate the wound healing assay, the sticker is removed to reveal the cell-free wound area. The plates are continually incubated in 5\% CO$_2$ at 37 $^{\circ}$C. The distribution of cells is imaged  at $t$ = 0, 9, 24, 33, 48, 57, 72 h for the assays initiated with the square and circular wound shapes.  The assays initiated with the triangular wounds are imaged at $t$ = 0, 9, 24, 33, 48, 57 h. For each wound shape we perform three identically prepared experimental replicates ($n=3$).

A proliferation assay is initiated in the same way as the wound healing assay except that there is no wound. The plates are continually incubated in 5\% CO$_2$ at 37 $^{\circ}$C and images are recorded at $t$ = 0, 9, 24, 33, 48, 57, 72, 81, 96 h. We perform one experimental replicate of the proliferation assay and analyse data from this assay by estimating the cell density in three different, identically--sized, rectangular subregions within the population (Johnston et al., 2015).

\begin{landscape}
\begin{figure}[p]
\centering
\includegraphics[width=1.5\textwidth]{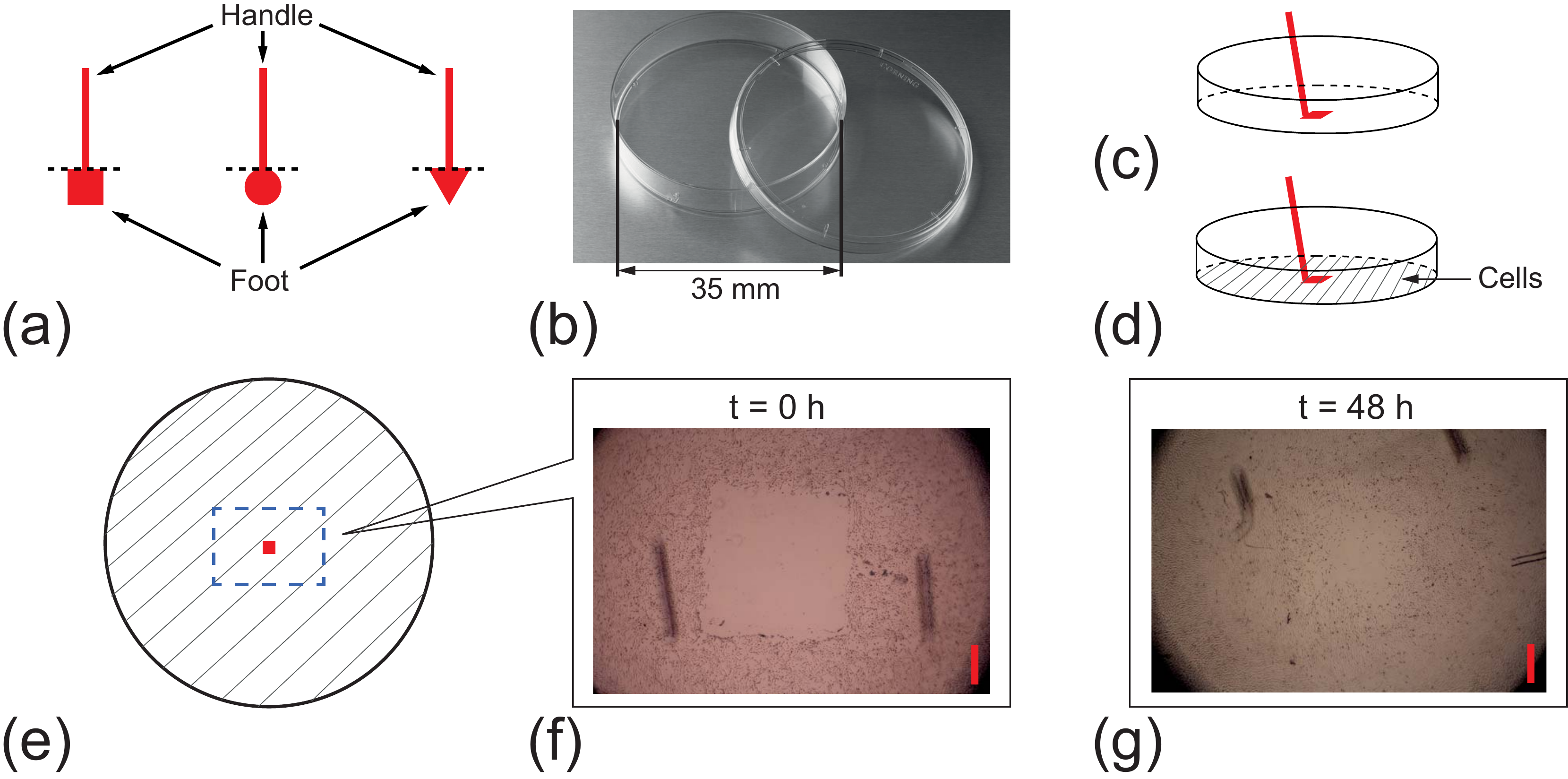}
\caption{{\bf Sticker assay protocol.}
(a) Stickers for creating square, circular, and triangular wound shapes. The side length and the diameter of square, equilateral triangle, and circular stickers are 2~mm. The red region shows both the foot of the sticker and the handle.  The dashed line indicates where the sticker is folded to create the handle so that the foot of the sticker can be easily attached, and removed from the tissue culture plate. (b) Corning$^\text{\textregistered}$ cell culture dish. (c) Isometric schematic of a 35~mm diameter cell culture plate showing the sticker attached to the plate before the cells are seeded into the dish. (d) Isometric schematic of the cell culture plate showing the apparatus after the cells are seeded into the dish.  (e) Top view of the experiment prior to the sticker being lifted. The blue dashed area shows the field of view that is imaged. (f) Experimental field of view at $t=0$ h with a square wound shape. (g) Experimental field of view at $t=48$ h showing the progression of the assay. The scale bar corresponds to 500 $\mu$m.}
\label{ExpAssay}
\end{figure}
\end{landscape}

\subsection{Edge detection method}\label{EdgeDetection}
We use ImageJ (Ferreira and Rasband, 2012) to detect the edges of the wound area in both the experimental and the simulation images (Treloar et al., 2013). For all images, the scale is first set using the \textit{Set Scale} function.  We find that 932 pixels corresponds to 1~mm for all experimental images, and 304 pixels corresponds to 1~mm for all the simulation images.  For each experimental image the contrast is enhanced by increasing the contrast and decreasing the brightness (\textit{Image}--\textit{Adjust}--\textit{Brightness/Contrast}).  The edges of the cells are then detected (\textit{Plugins}--\textit{Canny Edge Detector}), and enhanced using the Sobel method (\textit{Process}--\textit{Find Edges}).  Depending on the quality of the experimental image, the \textit{Find Edges} function may need to be used several times, both locally and globally for the edge of the wound area to be successfully detected. The edge of the wound area is automatically detected using the wand tracing tool. Then the wound area is calculated (\textit{Analyze}--\textit{Measure}). For each simulation image, the colour image is first set to grayscale (\textit{Image}--\textit{Type}--\textit{32-bit}). Then the edges of cells are detected (\textit{Plugins}--\textit{Canny Edge Detector}), and enhanced using the Sobel method (\textit{Process}--\textit{Find Edges}) two to three times. The wound area is automatically detected using the wand tracing tool and calculated (\textit{Analyze}--\textit{Measure}).  An important feature of our method is that we use the exact same image processing tools to quantify the time evolution of the area of the wound in both the experimental and the simulation images (Treloar et al., 2014).

\subsection{Mathematical methods}
\subsubsection{Discrete model}\label{DiscreteModel}
We use a discrete random walk model, in which each agent represents a single cell, to simulate the experiments. Simulations are performed on a hexagonal lattice, with the lattice spacing $\Delta$ that is taken to be equal to the average cell diameter of NIH 3T3 cells.  This gives $\Delta = 25$~$\mu$m (Treloar et al., 2013). Crowding effects are incorporated by ensuring that at most one agent can occupy a lattice site, and any potential motility events or proliferation events that would place more than one agent on a lattice site are aborted. Each lattice site, indexed $(i, j)$ where $i, j\in\mathbb Z^+$, has position
\begin{equation*}
(x, y) =
  \begin{cases}
    \left((i-1)\Delta, \sqrt{3}(j-1)\Delta/2\right) & \quad \text{if } j \text{ is even,}\\
    \left((i-1/2)\Delta, \sqrt{3}(j-1)\Delta/2\right) & \quad \text{if } j \text{ is odd,}\\
  \end{cases}
\end{equation*}
such that $1\leq i\leq I$ and $1\leq j\leq J$, so that $I$ and $J$ are chosen to accommodate the dimensions of the experimental field of view. In any single realisation of the random walk model, the occupancy of site $(i, j)$ is denoted $C_{i,j}$, with $C_{i,j} = 1$ if the site is occupied, and $C_{i,j} = 0$ if vacant.

If there are $N(t)$ agents present in the simulation at time $t$, then during the next time step of duration $\tau$, $N(t)$ agents are selected independently at random, one at a time with replacement, and given the opportunity to move (Simpson et al., 2010). The randomly selected agent attempts to move, with probability $P_m$, to one of the six nearest neighbour sites, with the target site chosen randomly. Motility events are aborted if an agent attempts to move to an occupied site. After the $N(t)$ potential motility events have been assessed, another $N(t)$ agents are selected independently, at random, one at a time with replacement, and given the opportunity to proliferate with probability $P_p$. The location of the daughter agent is chosen, at random, from one of the six nearest neighbour lattice sites. Potential proliferation events are aborted if the target site is occupied. However, if the target site is vacant, a new daughter agent is placed on that site. After the $N(t)$ potential proliferation events have been attempted, $N (t + \tau)$ is updated.

\subsubsection{Continuum limit of the discrete model}
While discrete models are useful to mimic and predict experimental observations, it is difficult to obtain more general insight using this approach.  Therefore, it is relevant to consider a mean field continuum limit description because we can then use additional mathematical and computational methods to gain insight into the model (O'Dea and King, 2012).
The mean field continuum limit description of the random walk model can be derived by formulating an approximate discrete conservation statement describing the change in average occupancy of site $\mathbf{s} = (i,j)$ during the interval from time $t$ to time $t + \tau$ (Simpson et al., 2010)
\begin{align}\label{ContinuumLimit}
\delta\langle C_{\mathbf{s}}\rangle= & +  \overbrace{\frac{P_m}{6}\left(1 - \langle C_{\mathbf{s}}\rangle\right)\sum_{\mathbf{s'} \in \mathcal{N}\{\mathbf{s}\}}\langle C_{\mathbf{s'}}\rangle}^{\text{change in occupancy due to migration into site} \ \mathbf{s}} \notag \\
&- \overbrace{\frac{P_m}{6}\langle C_{\mathbf{s}}\rangle \sum_{\mathbf{s'} \in \mathcal{N}\{\mathbf{s}\}}\left(1 - \langle C_{\mathbf{s'}}\rangle \right)}^{\text{change in occupancy due to migration out of site} \ \mathbf{s}} \nonumber\\ & + \overbrace{\frac{P_p}{6}\left(1 - \langle C_{\mathbf{s}}\rangle\right)\sum_{\mathbf{s'} \in \mathcal{N}\{\mathbf{s}\}}\langle C_{\mathbf{s'}}\rangle}^{\text{change in occupancy due to proliferation into site} \ \mathbf{s}},
\end{align}
where $\langle C_{\mathbf{s}}\rangle \in [0, 1]$ is the average occupancy of site $\mathbf{s}$, where the average is obtained by averaging the occupancy over a large number of identically prepared realisations, $\mathcal{N}\{\mathbf{s}\}$ is the set of six nearest-neighbour sites around site $\mathbf{s}$, and $\displaystyle{\sum_{\mathbf{s'} \in \mathcal{N}\{\mathbf{s}\}}\langle C_{\mathbf{s'}}\rangle}$ is the sum of the average occupancy of the nearest neighbour sites.  To proceed we invoke the usual mean field assumption which amounts to treating the average occupancy of lattice sites as independent (Simpson et al., 2010).

We then use Taylor series to expand each term in Eq. (\ref{ContinuumLimit}) about site $\mathbf{s}$ and neglect terms of $\mathcal{O}(\Delta^3)$. Dividing both sides of the resulting expression by $\tau$ and taking the limit as $\Delta \to 0$ and $\tau \to 0$ jointly, with the ratio $\Delta^2/\tau$ held constant, we identify $\langle C_{\mathbf{s}}\rangle$ with a smooth function, $C(x, y, t)$, that satisfies
\begin{equation}\label{PDE2d}
\dfrac{\partial C(x, y, t)}{\partial t} = \overbrace{D\nabla^2 C(x, y, t)}^{\text{unbiased motility mechanism with exclusion}} + \overbrace{r C(x, y, t) \left(1 - C(x, y, t)\right)}^{\text{unbiased proliferation mechanism with exclusion}},
\end{equation}
where $D$ [$\mu$m$^2$/h] is the cell diffusivity,
\begin{equation}\label{Diffusivity}
\displaystyle{D=\frac{P_m}{4}\lim_{\Delta \to 0, \tau\to 0} \left(\frac{\Delta^2}{\tau}\right)},
\end{equation}
and $r$ [/h] is the proliferation rate
\begin{equation}\label{ProliferationRate}
\displaystyle{r= \lim_{\Delta \to 0, \tau\to 0} \left(\frac{P_p}{\tau}\right)}.
\end{equation}

Therefore, the continuum limit description of this discrete model is the two-dimensional analogue of the well-known Fisher--Kolmorogov model, which has been used previously to study wound healing experiments (Ascione et al., 2017; Sheardown and Cheng, 1996;  Tremel et al. 2009).  Instead of working with a continuum model alone, we find that it is useful to work with both the discrete random walk model and the continuum limit description because this gives us a better opportunity to describe certain features of the experiments rather than working with just the continuum model in isolation.

Note that the maximum carrying capacity density in the discrete model is unity, since the maximum number of agents per lattice site is one.  Similarly, the carrying capacity density in Eq. (\ref{PDE2d}) is also unity.  However, the carrying capacity density in the experiments will take on some positive value, $K$.  Therefore, to apply our model to match the dimensional experiments we re-dimension the dependent variable, $\mathcal{C}(x, y, t) = C(x, y, t) K$, where $K$ [cells/$\mu$m$^2$] is the dimensional carrying capacity density.  In dimensional variables, Eq. (\ref{PDE2d}) can be written as
\begin{equation}\label{PDE2d-2}
\dfrac{\partial \mathcal{C}(x, y, t)}{\partial t} =D\nabla^2 \mathcal{C}(x, y, t) + r \mathcal{C}(x, y, t) \left(1 - \frac{\mathcal{C}(x, y, t)}{K}\right).
\end{equation}
We will use the dimensional continuum limit model to match data obtained from experimental images.

\subsubsection{Simplified continuum model for cell proliferation assays}
When modelling the proliferation assays in which there is, on average, no spatial gradient of cell density we can simplify Eq. (\ref{PDE2d-2}) since $\nabla^2 \mathcal{C}(x, y, t) = 0$ in these experiments. Therefore, the two-dimensional PDE, with independent variable $\mathcal{C}(x, y, t)$, simplifies to an ordinary differential equation (ODE) with independent variable $\mathcal{C}(t)$ that is given by
\begin{equation}\label{ODE}
 \frac{\textrm{d}\mathcal{C}(t)}{\textrm{d}t} = r \mathcal{C}(t)\left(1 - \frac{\mathcal{C}(t)}{K}\right),
\end{equation}
where $\mathcal{C}(t)$ [cells/$\mu$m$^2$] is the dimensional cell density, and $t$ [h] is time. Equation (\ref{ODE}) is the logistic growth model, and the solution is
\begin{equation}\label{LogisticGrowthExactSoln}
\mathcal{C}(t) = \frac{K \mathcal{C}(0)}{\left(K - \mathcal{C}(0)\right)\mathrm{e}^{-r t} + \mathcal{C}(0)}.
\end{equation}

\subsection{Motivation}
While it is obvious that real wounds take on arbitrary shapes and sizes, \textit{in vitro} wound healing assays are almost always limited to just one particular wound shape. Therefore, the role of initial wound shape in \textit{in vitro} experimental models of wound healing is poorly understood because it has not been previously examined. An implicit assumption, that is rarely stated and never tested, is that when an \textit{in vitro} wound healing assay with a specific initial wound shape is performed, the results could be extrapolated to apply to a different situation where a wound is created with a different shape. For example, a relevant question for us to consider is if we perform a sticker assay with a square wound, can the results from that assay be applied to predict the closure of a circular wound?  This question motivates our present work in which we perform, and analyse, a series of sticker assays with a range of initial wound shapes.  Using our mathematical model we calibrate values of $r$, $K$ and $D$ so that our mathematical model matches the observations from the experimental data, and we can quantitatively assess the role of wound shape by comparing parameter estimates obtained by considering experiments with different initial wound shape.  We are particularly interested in this question because recent studies that combine \textit{in vitro} wound healing assays with mathematical models show that the results of these experiments can be extremely sensitive to the initial configuration of cell in the experiments (Jin et al., 2016; Treloar et al., 2014).

\section{Results and discussion}
\subsection{Estimating the rate of wound closure}\label{Estimatingw}
\begin{landscape}
\begin{figure}[p]
\centering
\includegraphics[width=1.5\textwidth]{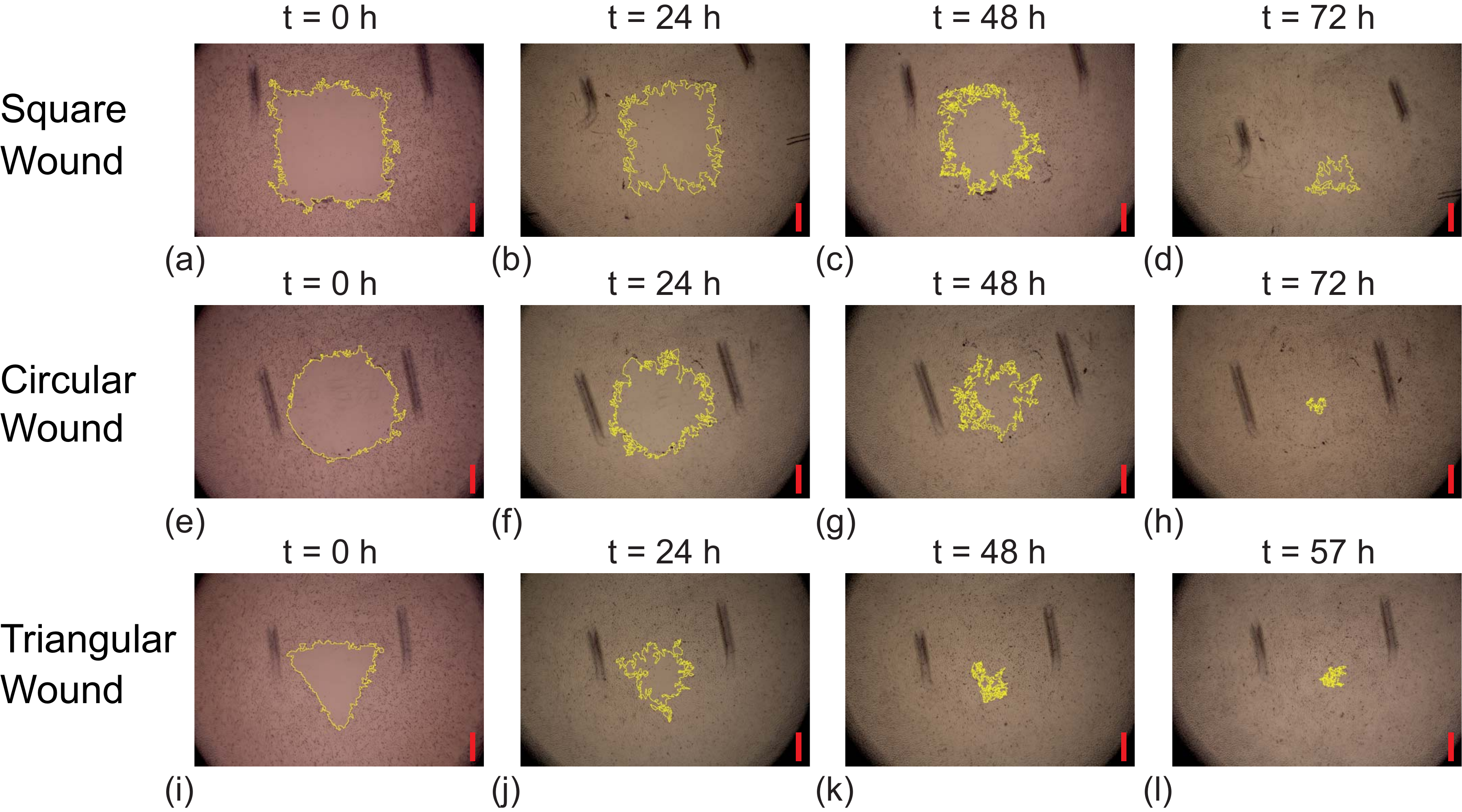}
\caption{{\bf Experimental images superimposed with the position of the leading edge.} Images of wound healing assays at $t = 0, 24, 48$, and 72 h with initially (a)--(d) square, (e)--(h) circular, and (i)--(l) triangular wound shapes. The detected edges are highlighted with the yellow colour. The scale bar corresponds to 500 $\mu$m.}
\label{woundshapeexp}
\end{figure}
\end{landscape}

We use edge detection methods to locate the position of the leading edge of the wound and to calculate the wound area in all experimental images.   To provide some qualitative comparison of the detected leading edge we superimpose the leading edges on the experimental images in Fig. \ref{woundshapeexp}.  Visual interpretation of the position of the detected edges suggests that the edge detection algorithm  clearly and accurately detects the edge of the populations in all assays we consider.    Results in Fig. \ref{woundshapeexp} compare the experimental images and the position of the detected leading edge for one of the experimental replicates only.  Similar data, showing a visual comparison of the experimental images and the position of the detected leading edge for the remaining experimental replicates are given in the Supplementary Material document.

\begin{landscape}
\begin{figure}[p]
\centering
\includegraphics[width=1.5\textwidth]{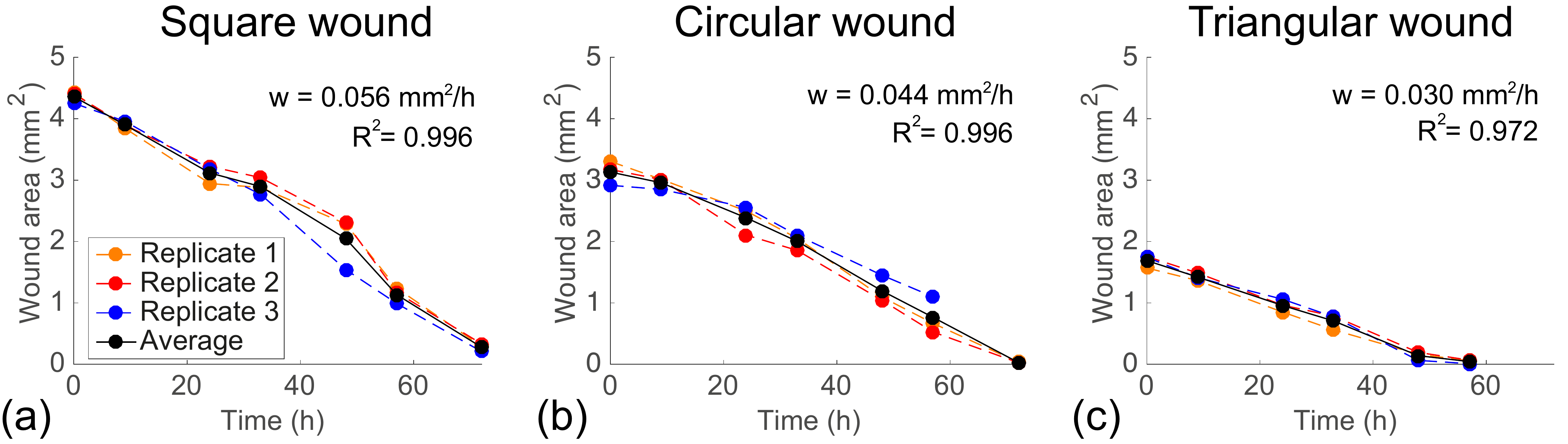}
\caption{{\bf Time evolution of the wound area.} The experimental wound area for, (a) square, (b) circular, and (c) triangular wound shape is given. The wound data is given for each experimental replicate as well as the averaged wound area. The rate of wound closure, $w$, and the coefficient of determination, $R^2$, are indicated.}
\label{woundshapeexp2}
\end{figure}
\end{landscape}

The data in Fig. \ref{woundshapeexp} allows us to quantify the progression of the experiments by measuring the area enclosed by the detected edge, and examining how this area decreases with time as the wound closes.  We plot the time evolution of the wound area as a function of time for each initial wound shape in Fig. \ref{woundshapeexp2}. Visual analysis of this data suggests that the wound area decreases approximately linearly for all initial wound shapes.  To quantify the rate of closure we fit a straight line to the averaged data in Fig. \ref{woundshapeexp2}.  The slope of the linear least--squares line, $w$, gives us a measure of the rate of wound closure, and we find that $w= 0.056, 0.044$, and 0.030 mm$^2$/h for the square, circular and triangular wounds, respectively.

Interestingly, this data suggests that the wound closure rate varies, with the wound closure rate for the square wound almost twice the wound closure rate for the triangle.  This is an intriguing result. Presenting data in this way is a standard approach, and we might have anticipated that since the only difference in the experiments is the initial wound shape that we might see very little differences in the rate of wound closure.  We hypothesize that this difference could have two possible explanations:

\begin{enumerate} \itemsep=5mm
\item  Perhaps cells behave differently (i.e. have different rates of motility and/or proliferation) when they are subjected to different initial wound shapes;
\item Perhaps the differences in wound closure rates occur directly as a result of the differences in initial wound shape, and there is no difference in the underlying behaviour of individual cells.
\end{enumerate}

To make a distinction between these two potential explanations, we now calibrate the mathematical model to our experimental data to provide estimates of the cell proliferation rate, carrying capacity density, and cell diffusivity for each initial wound shape.  If we find that the parameter values depend on the initial geometry then it might be reasonable to conclude that the initial geometry directly influence the behaviour of cells.  In contrast, if the parameter values do not depend on the initial wound geometry then it would be reasonable to conclude that the initial geometry plays no direct role on the fundamental cell behaviour.

\subsection{Estimating the cell proliferation rate and carrying capacity density}\label{Estimatingrnk}
\begin{landscape}
\begin{figure}[p]
\centering
\includegraphics[width=1.5\textwidth]{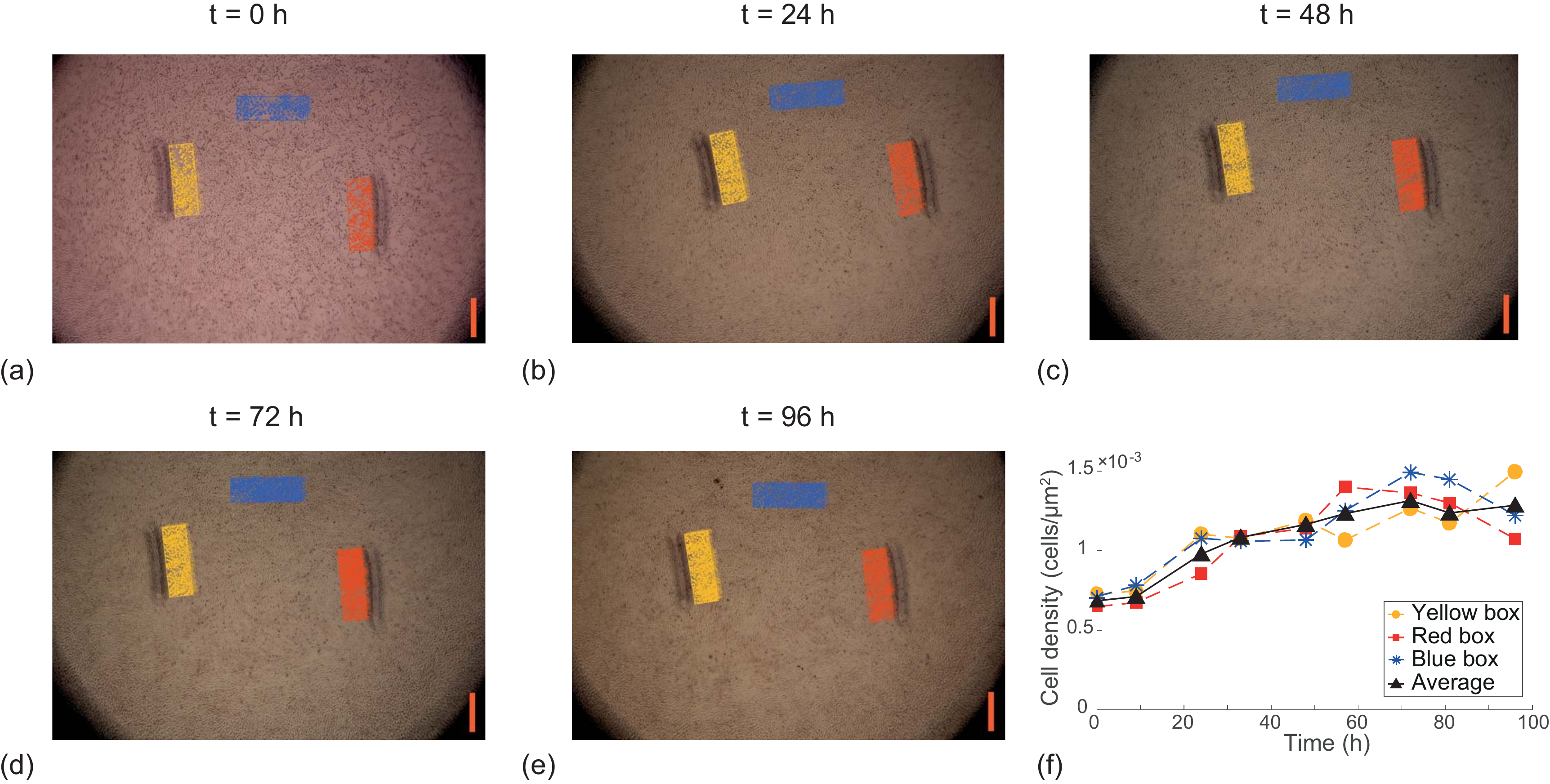}
\caption{{\bf Estimating $r$ and $K$.}
(a) - (e) Series of images showing the progression of the proliferation assays. The time at which the image is recorded is indicated, and the scale bar corresponds to 500 $\mu$m. In each subfigure, three 900 $\mu$m $\times$ 300 $\mu$m subregions, highlighted with yellow, red and blue rectangles, respectively, are superimposed on the experimental image. Manual cell counting is used to estimate the number of cells in each subregion. (f) Cell density information is obtained at $t$ = 0, 9, 24, 33, 48, 57, 72, 81 and 96 h.}
\label{cellcount}
\end{figure}
\end{landscape}

We count the number of cells in three identically-sized rectangular subregions in the cell proliferation assays, as highlighted in Fig. \ref{cellcount}.  Each subregion has dimensions 900 $\mu$m $\times$ 300 $\mu$m, and the total number of cells in each subregion are counted in Photoshop using the `Count Tool' (Adobe Systems Incorporated, 2017). After counting the number of cells in each subregion, we divide the total number of cells by the total area to estimate the cell density at $t$ = 0, 9, 24, 33, 48, 57, 72, 81 and 96 h.  All raw data are included in the Supplementary Material document.

\begin{landscape}
\begin{figure}[p]
\centering
\includegraphics[width=1.6\textwidth]{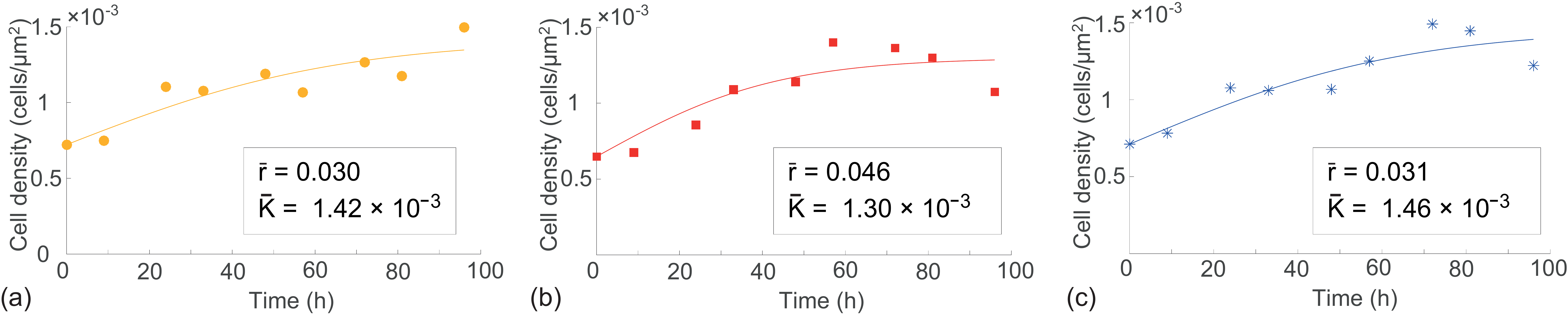}
\caption{{\bf Comparing the mathematical model prediction and experimental data for the proliferation assay.}
The solution of the logistic growth model is calibrated to the data of cell density information in (a) yellow, (b) red, and (c) blue boxes shown in Fig \ref{cellcount}. In each subfigure, the solid line represents the calibrated solution, and the individual markers represent the experimental data. The least--squares estimates of $r$ and $K$ are shown in the text box.}
\label{rnk}
\end{figure}
\end{landscape}

Using the data from the cell proliferation assay, we calibrate the solution of the logistic growth model to the cell density information in all three subregions, as shown in Fig. \ref{cellcount}(f).  This procedure allows us to estimate the cell proliferation rate $r$, and the carrying capacity density $K$. To calibrate the model we form a least--squares measure of the discrepancy between the solution of the logistic growth model and the cell density data in each of the three subregions. This least--squares measure is given by
\begin{equation}
E(r, K) = \sum_{l=1}^{8}\left[ \mathcal{C}^{\textrm{model}}(t_l) - \mathcal{C}^{\textrm{data}}(t_l) \right]^2,
\end{equation}
where $l$ is an index that indicates the number of time points. To find values of $r$ and $K$ that minimise $E(r, K)$, we use the MATLAB function \textit{lsqcurvefit} (MathWorks, 2017) that is based on the Levenberg--Marquardt algorithm.  We always take care to ensure that the iterative method is insensitive to our initial choice of $r$ and $K$.  We denote the minimum least--squares error as $E_{\text{min}} = E (\bar{r}, \bar{K})$.  To calibrate the logistic growth model to data from the proliferation assay, we use data $t=0$ in Fig. \ref{cellcount}(f) as the initial condition in Eq. (\ref{LogisticGrowthExactSoln}), and match Eq. (\ref{LogisticGrowthExactSoln}) to the rest of the data.  We repeat this procedure three times using experimental data from each of the three subregions.  To demonstrate the quality of the match between the experimental data and the calibrated logistic growth model, we superimpose the experimental data and Eq. (\ref{LogisticGrowthExactSoln}) with the estimates of $\bar{r}$ and $\bar{K}$ for each subregion in Fig. \ref{rnk}. The results indicate that the quality of match between the solution of the logistic model and the experimental data is very good. Estimates of $\bar{r}$ and $\bar{K}$ for each subregion are summarised in Table \ref{rnkTable}.   Since the variation in  $\bar{r}$ and $\bar{K}$ between the three subregions is relatively small, we further average these estimates to give overall estimates of  $r = 0.036$ /h and $K = 1.4 \times 10^{-3}$ cells/$\mu$m$^2$.  We note that these estimates of $r$ and $K$ are consistent with previous estimates for 3T3 fibroblast cells (ATCC, 2017; Treloar et al., 2014).

\begin{landscape}
\vspace{1cm}
\begin{table}[ht]
\caption{Estimates of $\bar{r}$ and $\bar{K}$ for the logistic growth model obtained by calibrating Eq. (\ref{LogisticGrowthExactSoln}) to the average cell density information for the three subregions. The third column gives the doubling time, $t_d = \textrm{ln}(2)/\bar{r}$. All parameter estimates are given to two significant figures.}
\renewcommand{\arraystretch}{1.25}
\centering
\begin{tabular}{ | c | c | c | c | c |}
\hline
Subregion & $\bar{r}$ (/h) & $t_d$ (h)  & $\bar{K}$ (cells/$\mu$m$^2$) & $E_{\text{min}}$ ((cells/$\mu$m$^2$)$^2$)\\ [0.5ex]
\hline
Yellow & 0.030 & 23 & 1.4 $\times$ 10$^{-3}$ & 8.7 $\times$ 10$^{-8}$ \\
\hline
Red & 0.046 & 15 & 1.3 $\times$ 10$^{-3}$ & 1.2 $\times$ 10$^{-7}$ \\
\hline
Blue & 0.031  & 22 & 1.5 $\times$ 10$^{-3}$ & 9.4 $\times$ 10$^{-8}$ \\
\hline
Mean $\pm$ Standard deviation & 0.036 $\pm$ 0.009 & 19 $\pm$ 4 & 1.4 $\pm$ 0.07 $\times$ 10$^{-3}$ & -\\
\hline
\end{tabular}
\label{rnkTable}
\end{table}
\end{landscape}

\subsection{Estimating the cell diffusivity}\label{EstimatingD}
We now apply the discrete random walk model to mimic the sticker assays to estimate the cell diffusivity. Our estimate of $r$ in Section \ref{Estimatingrnk} allows us to specify the ratio $P_p/\tau$.  We note that the individual values of $P_p$ and $\tau$ are not uniquely specified, but if we take $\tau = 0.05$ h and  $P_p = 0.0018$, then the proliferation rate in the discrete model gives $r = P_p/\tau = 0.036$ /h, as previously obtained.  To estimate $D$, we vary the parameters in the model to examine the behaviour of the model when we consider the diffusivity parameter to lie within the interval, $500 \le D \le 2000$ $\mu$m$^2$/h.  We choose this interval since the cell diffusivity for fibroblast cells is thought to lie within this range (Johnston et al. 2016).  Since we have $\Delta = 25$ $\mu m$ (Treloar et al., 2014), this interval of $D$ corresponds to an interval of $0.16 \le P_m \le 0.64$, and we seek to find a value of $P_m$, and hence $D$, which provides the best match between the discrete model and the experimental images for each wound shape.

To simulate the sticker assays shown in Fig. \ref{ExpAssay}, we use a lattice of size 10~mm $\times$ 10~mm, which can be accommodated by setting $I = 401$ and $J = 463$ with $\Delta = 25$ $\mu$m.  This simulation lattice is much smaller than the total size of the domain, which is a circular tissue culture plate of diameter 35~mm.  However, the simulation lattice is considerably larger than the experimental field of view, which is 5.56~mm $\times$ 3.71~mm, as shown in Fig. \ref{ExpAssay}.  Since the field of view is smaller than the simulation domain, and cells in the experiment are distributed uniformly away from the initial placement of the sticker, there will be zero net flux of cells across the boundary of the field of view for all time (Johnston et al., 2015).   Therefore, specifying zero net flux conditions on the boundary of the simulation lattice will mimic these experimental conditions.  To initialise the random walk simulations we note that the initial cell density in the proliferation assays is approximately 40\% of the carrying capacity density.  Therefore, we initialise the random walk simulations of the sticker assays by randomly occupying each lattice site with probability 40\%.  To simulate how the presence of the sticker prevents cells from occupying certain regions in the experiment, we then we remove all agents within an appropriately sized square, circle or equilateral triangle in the centre of the simulation lattice.

\begin{landscape}
\begin{figure}[h]
\centering
\includegraphics[width=1.4\textwidth]{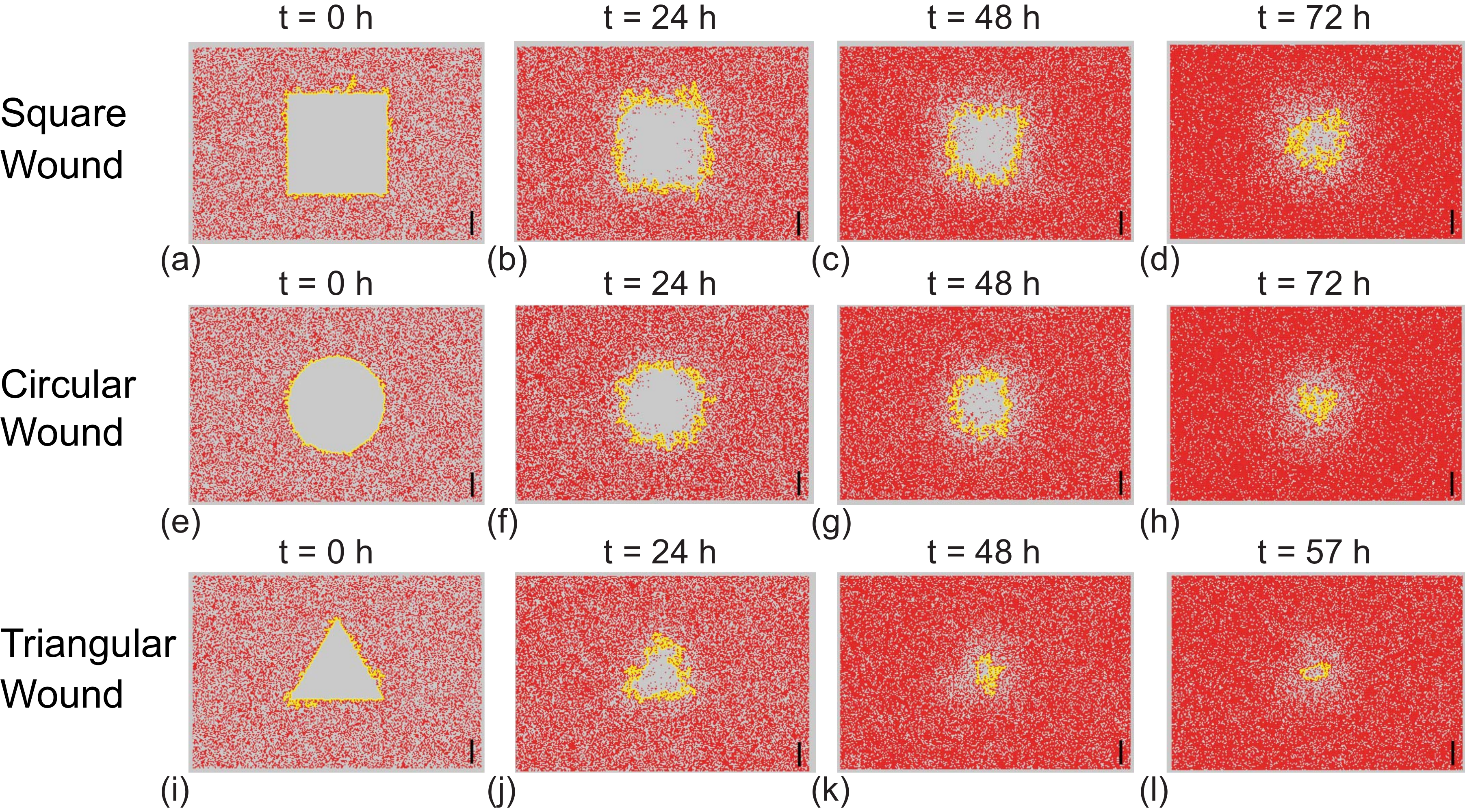}
\caption{{\bf Random walk simulation of sticker assays with the leading edge superimposed.} Images of simulation data for sticker assays at $t = 0, 24, 48, 72$ h with initially (a)--(d) square, (e)--(h) circular, and (i)--(l) triangular wound shapes. The detected leading edges (yellow) are superimposed. Each simulation is initiated by randomly populating a lattice with probability 40\%. A square, circular, and triangular wound is made at $t=0$ h in (a), (e), and (i), respectively. All simulations correspond to $\Delta = 25$ $\mu$m, $\tau = 0.05$ h, $P_p = 0.0018$ and $P_m=0.32$. The scale bar corresponds to 500 $\mu$m.}
\label{woundshapesimu}
\end{figure}
\end{landscape}

\begin{landscape}
\begin{figure}[h]
\centering
\includegraphics[width=1.4\textwidth]{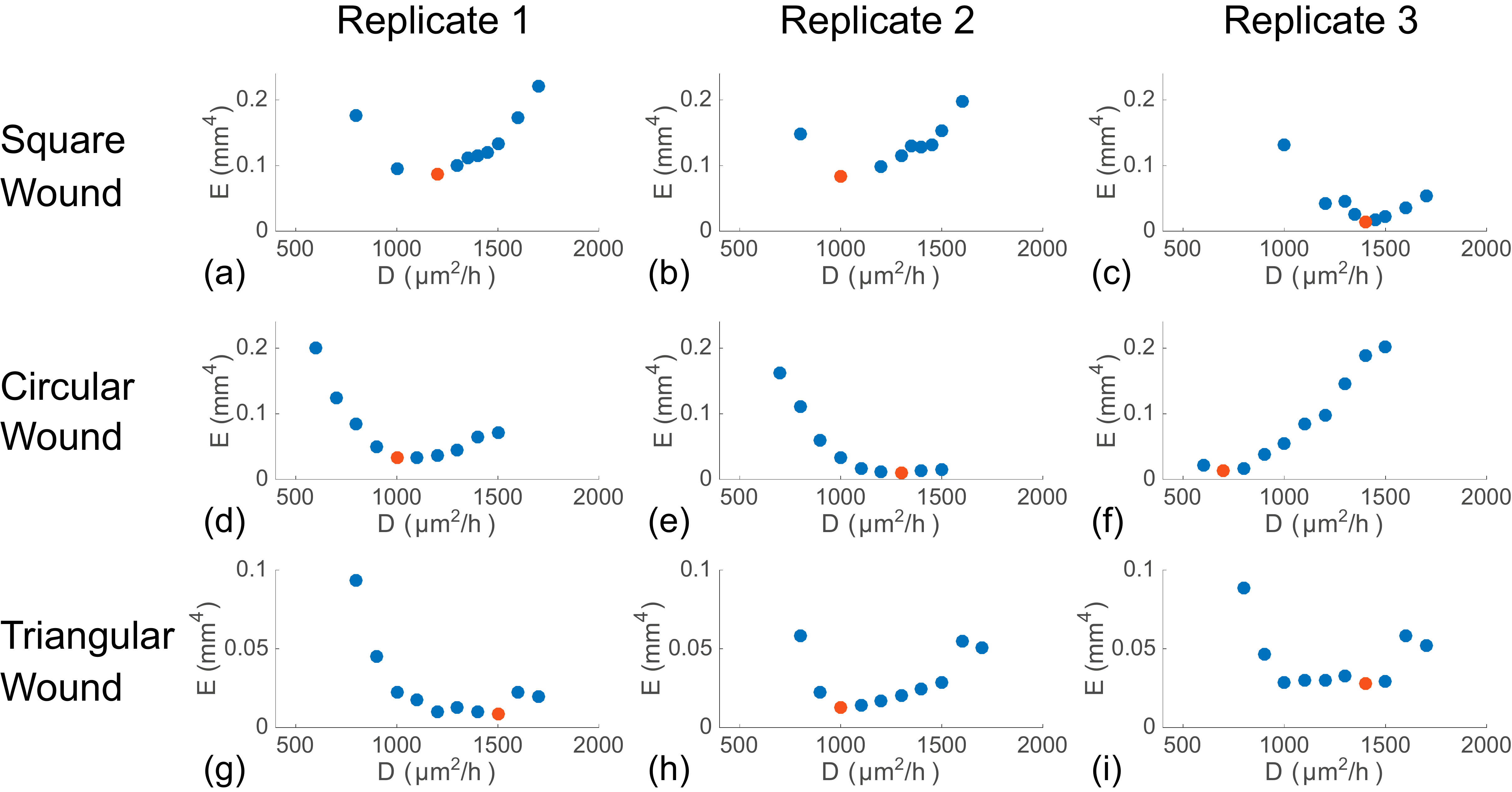}
\caption{{\bf Least--squares error to determine the cell diffusivity.}
(a)--(c) The least--squares error for the square wound shape. The average $\bar{D}$ over the three replicates is 1200 $\pm$ 260 $\mu$m$^2$/h. (d)--(f) The least--squares error for the circular wound shape. The average $\bar{D}$ over the three replicates is 1000 $\pm$ 300 $\mu$m$^2$/h. (g)--(i) The least--squares error for the triangular wound shape. The average $\bar{D}$ over the three replicates is 1300 $\pm$ 265 $\mu$m$^2$/h. In each subfigure the red circle represents the estimate of $\bar{D}$ that minimises the least--squares error. All simulations correspond to $\Delta = 25$ $\mu$m, $\tau = 0.05$ h, and $P_p = 0.0018$.}
\label{leastsquares}
\end{figure}
\end{landscape}

Figure \ref{woundshapesimu} shows representative snapshots from the discrete model initialised with the three different initial wound shapes.  Although the simulation lattice is larger than the experimental field of view, we present our results from the discrete model by showing a region of the lattice that has the same dimensions of the experimental field of view, as shown in Fig. \ref{woundshapeexp}. The snapshots of the simulated wound closing for each initial wound shape show similar qualitative trends to those in the experimental images. To estimate $D$, we systematically vary $P_m$ in the simulations.  For each value of $P_m$ we generate a series of snapshots from the discrete model and use the ImageJ edge detection method to find the location of the wound edge in each simulation snapshot. We estimate $D$ by examining a measure of the difference between the area enclosed by the leading edge in the experimental images and averaged data from three identically prepared stochastic simulations.  The measurement of discrepancy we consider is given by
\begin{equation}
E(D) = \frac{1}{L}\sum_{l=1}^{L}\left[ A^{\textrm{model}}(t_l) - A^{\textrm{data}}(t_l) \right]^2,
\end{equation}
where $A^{\textrm{model}}(t_l)$ is the average area of the wound estimated from the discrete mathematical model at time $t_l$, and $A^{\textrm{data}}(t_l)$ is the area of the wound estimated from the experimental image at time $t_l$.  Here,  $l = 1,2, \ldots, L$ is an index indicating the number of time points used to compare the experimental images with images generated from the random walk model.  Results in Fig. \ref{leastsquares} show $E(D)$ for the three wound shapes, and we identify $\bar{D}$ as the value of $D$ that minimises the discrepancy, $E_{\text{min}} = E(\bar{D)}$.  Our estimates of $\bar{D}$ are summarised in Table \ref{DTable}.  There are two notable features of these estimates: (1) The difference between the average $\bar{D}$ for the three wound shapes is relatively small; and (2) the range of estimated $\bar{D}$ for the three wound shapes overlap. Therefore, the simplest explanation of our model calibration procedure is that our estimates of $D$ are effectively independent of the initial wound shape.

\begin{table}[ht]
\caption{Estimates of $\bar{D}$.  All parameter estimates are given to two significant figures.}
\renewcommand{\arraystretch}{1.25}
\centering
\begin{tabular}{ | c | c | c |}
\hline
Square wound shape  & $\bar{D}$ ($\mu$m$^2$/h) & $E_{\text{min}}$ (mm$^4$)\\ [0.5ex]
\hline
Experimental replicate 1 & 1200 & 8.6 $\times$ 10$^{-2}$ \\
\hline
Experimental replicate 2 & 1000 & 8.4 $\times$ 10$^{-2}$ \\
\hline
Experimental replicate 3 & 1400 & 1.5 $\times$ 10$^{-2}$ \\
\hline
Mean $\pm$ Standard deviation & 1200 $\pm$ 200 & - \\
\hline
\hline
Circular wound shape  & $\bar{D}$ ($\mu$m$^2$/h) & $E_{\text{min}}$ (mm$^4$)\\ [0.5ex]
\hline
Experimental replicate 1 & 1000 & 3.3 $\times$ 10$^{-2}$ \\
\hline
Experimental replicate 2 & 1300 & 1.0 $\times$ 10$^{-2}$ \\
\hline
Experimental replicate 3 & 700 & 1.3 $\times$ 10$^{-2}$ \\
\hline
Mean $\pm$ Standard deviation & 1000 $\pm$ 300 & - \\
\hline
\hline
Triangular wound shape  & $\bar{D}$ ($\mu$m$^2$/h) & $E_{\text{min}}$ (mm$^4$)\\ [0.5ex]
\hline
Experimental replicate 1 & 1500 & 8.0 $\times$ 10$^{-3}$ \\
\hline
Experimental replicate 2 & 1000 & 1.2 $\times$ 10$^{-2}$ \\
\hline
Experimental replicate 3 & 1400 & 2.8 $\times$ 10$^{-2}$ \\
\hline
Mean $\pm$ Standard deviation & 1300 $\pm$ 270 & - \\
\hline
\end{tabular}
\label{DTable}
\end{table}

Therefore, if we pool all of these estimates and work with two significant figures only, our overall estimate of the cell diffusivity is $1200 \pm 260$ $\mu$m$^2$/h.

To provide an additional check on the ability of our mathematical model to predict the temporal evolution of the experiments, we will now explore how well the solution of Eq. (\ref{PDE2d-2}), parameterised with $r = 0.036$ /h, $K = 1.4 \times 10^{-3}$ cells/$\mu$m$^2$, and $D = 1200$ $\mu$m$^2$/h, provides a useful prediction of the experimental data.
To do this we solve Eq. (\ref{PDE2d-2}) on the same domain as we use in our previous discrete simulations, and we apply zero net flux boundary condition on all boundaries. The initial condition is given by $\mathcal{C}(x,y,0) = 0$ within the initial wound area, and $\mathcal{C}(x,y,0) = 0.4K$ outside the wound area. Using the method of lines, Eq. (\ref{PDE2d-2}) is discretised with a central difference approximation with uniform node spacing, $\delta$.  Details of the discretisation are provided in the Supplementary Material document. The resulting system of coupled nonlinear ordinary differential equations is solved using MATLAB function \textit{ode45} with tolerance $\epsilon$ (MathWorks, 2017).  We superimpose the numerical solution of Eq. (\ref{PDE2d-2}) on the experimental images by showing the contour, $\mathcal{C}(x,y,t) = 0.2K$, at various times.  This contour corresponds to a density of 20\% of the confluent density, and we find that this choice of contour provides a good estimate of the density at the leading edge as detected by the ImageJ edge detection algorithm. Details of the procedure used to provide this estimate are given in the Supplementary Material document.  Overall, comparing the solution of Eq. (\ref{PDE2d-2}) presented in terms of the contour of the leading edge density and the experimental images suggests that our mathematical model, parameterised with a unique combination of parameters, can predict the time evolution of the wound area for all three  initial wound shapes.

\begin{landscape}
\begin{figure}[p]
\centering
\includegraphics[width=1.4\textwidth]{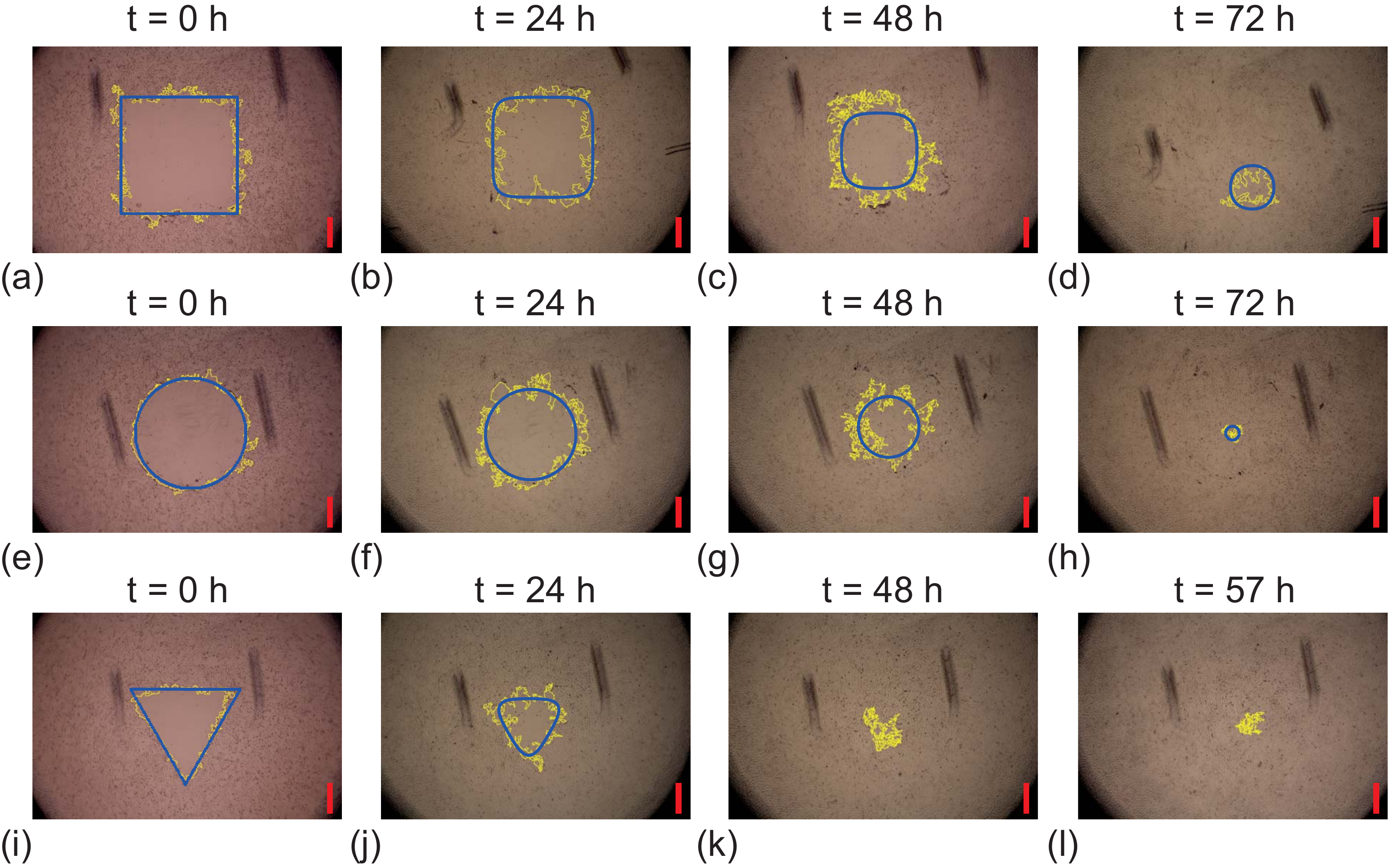}
\caption{{\bf Contour plot of numerical solutions of Eq. (\ref{PDE2d}) for the sticker assays.}
The wound closure with initially square, circular and triangular wound shapes are shown in (a)--(d), (e)--(h), and (i)--(l), respectively. The blue contour plot that corresponds to $\mathcal{C}(x, y, t) = 0.2K$ is superimposed onto the detected edge in each subfigure. The scale bar corresponds to 500 $\mu$m. The numerical solutions of  Eq. (\ref{PDE2d-2}) is obtained with $D = 1200$ $\mu$m$^2$/h, $r = 0.036$ /h, $K = 1.4 \times 10^{-3}$ cells/$\mu$m$^2$, $\delta  = 10$ $\mu$m, and $\epsilon = 1 \times 10^{-5}$.}
\label{Contour}
\end{figure}
\end{landscape}

\section{Conclusions}
Wound healing assays are routinely used to study the collective cell migration during wound closing (Keese et al., 2004; Singh et al., 2003). While real wounds take on arbitrary shapes, wound healing assays are usually limited to just one particular wound shape (Gough et al., 2011; Jin et al., 2016; Johnston et al., 2015, 2016; Keese et al., 2004; Kramer et al., 2013; Riahi et al., 2012; Sengers et al., 2007).  When we interpret results from a wound healing assay, an implicit assumption is always made.  This assumption is that the results from a particular assay, with a  specific initial wound shape, would apply to other wounds with a different initial shape. This implicit assumption is always made,  rarely stated and never examined in any detail. To explore the validity of such an assumption, here we develop and perform a new kind of wound healing assay, called a \textit{sticker assay}, to examine wound healing with various initial wound shapes.

Previous experimental studies present the results from wound healing assays by reporting the time evolution of wound area as the wound closes (Leu et al., 2012; Ueck et al., 2017; Yarrow et al., 2004).   When we report the results from our sticker assays in this standard way, we find that the area of the circular, square and triangular wounds close linearly with time.  However, the rate of wound closure is very different between the three initial wound shapes.  Without further examination, this kind of standard data might suggest that the mechanisms driving wound closure could depend on the initial wound shape.  To provide further information about this question we attempt to quantify the relevant mechanisms in the experiments by calibrating the solution of a discrete random walk model, and the continuum-limit description of this model, to the experimental data.

In summary, we find that our estimates of the cell diffusivity for each initial wound area are similar, and the range of cell diffusivity obtained from the three experimental replicates for each initial wound shape overlap.  Therefore, the simplest possible explanation of our results is that the two-dimensional Fisher-Kolmogorov model with one unique choice of parameters provides a good match of the experimental data.  Therefore, while the temporal wound area data  depends on the initial wound shape, the underlying mechanisms that drive the behaviour of the cell populations (i.e. cell proliferation and cell migration) do not depend on the initial wound shape.

To provide a confirmation of our results, we solve the two--dimensional Fisher-Kolmogorov equation with our single set of parameter values for each initial wound shape. To check that the model matches the experimental data we superimpose a particular contour from the numerical solution with $\mathcal{C}(x,y,t) = 0.20K$ onto the experimental images. This comparison implies that the continuum-limit PDE description of our random walk model, parameterised with a unique combination of parameters, provides a good match to the experimental data.  Again, this result implies that while the temporal wound area data depends on the initial wound shape, the fundamental transport mechanism that drive the wound healing processes not to depend on the initial wound shape.

Our approach to modelling experimental data is always to use the simplest possible mathematical model that describes the key features in the experiment.  In this case we use a discrete exclusion process in which agents undergo unbiased migration and unbiased proliferation.  The unbiased exclusion process-based motility mechanism gives rise to a linear diffusion term in the continuum limit PDE, and the exclusion process-based proliferation mechanism gives rise to a logistic source term.  Therefore, the continuum limit PDE is the two-dimensional Fisher-Kolmorogov model, and our data suggests that this model provides a good match to the experimental observations.  However, we are well aware that other studies suggest that the Porous--Fisher model, which has a nonlinear diffusion term, might be preferable since this model gives rise to well-defined sharp fronts  (Maini et al., 2004a, 2004b; Sengers et al., 2007). Since we find that the simpler model with a linear diffusion continuum-limit provides a good match to the experimental data, we do not pursue using any kind of more complicated mathematical model at this stage.

From a mathematical perspective, the geometry of our wound healing assay is reminiscent of a \textit{hole-closing problem}.  These problems are characterised by partial differential equations being applied outside of a two-dimensional ``hole'' which shrinks inwards in time.  Typically, these are formulated as moving boundary problems with Stefan-type boundary conditions (Dallaston and McCue, 2013; McCue and King, 2011) or, alternatively, nonlinear diffusion problems with degenerate diffusive terms and sharp interfaces (Angenent et al., 2001; Betelu et al., 2000; Witelski, 1995).  As an alternative to using a discrete random walk model, it would be interesting to model our experimental data as a hole-closing problem and explore the effects of cell migration and proliferation on the geometry of the wound, as predicted by that model, as the wound closes.


\end{document}


\begin{frontmatter}

\title{Supplementary Material}
\author{Wang Jin$^{1}$, Kai-Yin Lo$^{2}$, Shih--En Chou$^{2}$, }
\author{Scott W McCue$^{1}$}
\author{$^*$Matthew J Simpson$^{1}$}
\corauth[cor]{Corresponding author}
\address{$^1$ School of Mathematical Sciences, Queensland University of Technology (QUT) Brisbane, Queensland 4000, Australia.}
\address{$^2$ Department of Agricultural Chemistry, National Taiwan University\\
Taipei 10617, Taiwan.}
\ead{matthew.simpson@qut.edu.au, \textit{\textrm{Telephone}} +
617 31385241, \textit{\textrm{Fax}} + 617 3138 2310}

\end{frontmatter}

\newpage
\section{Experimental data: Wound area}
\begin{landscape}
\begin{figure}
\begin{center}
\includegraphics[width=1.6\textwidth]{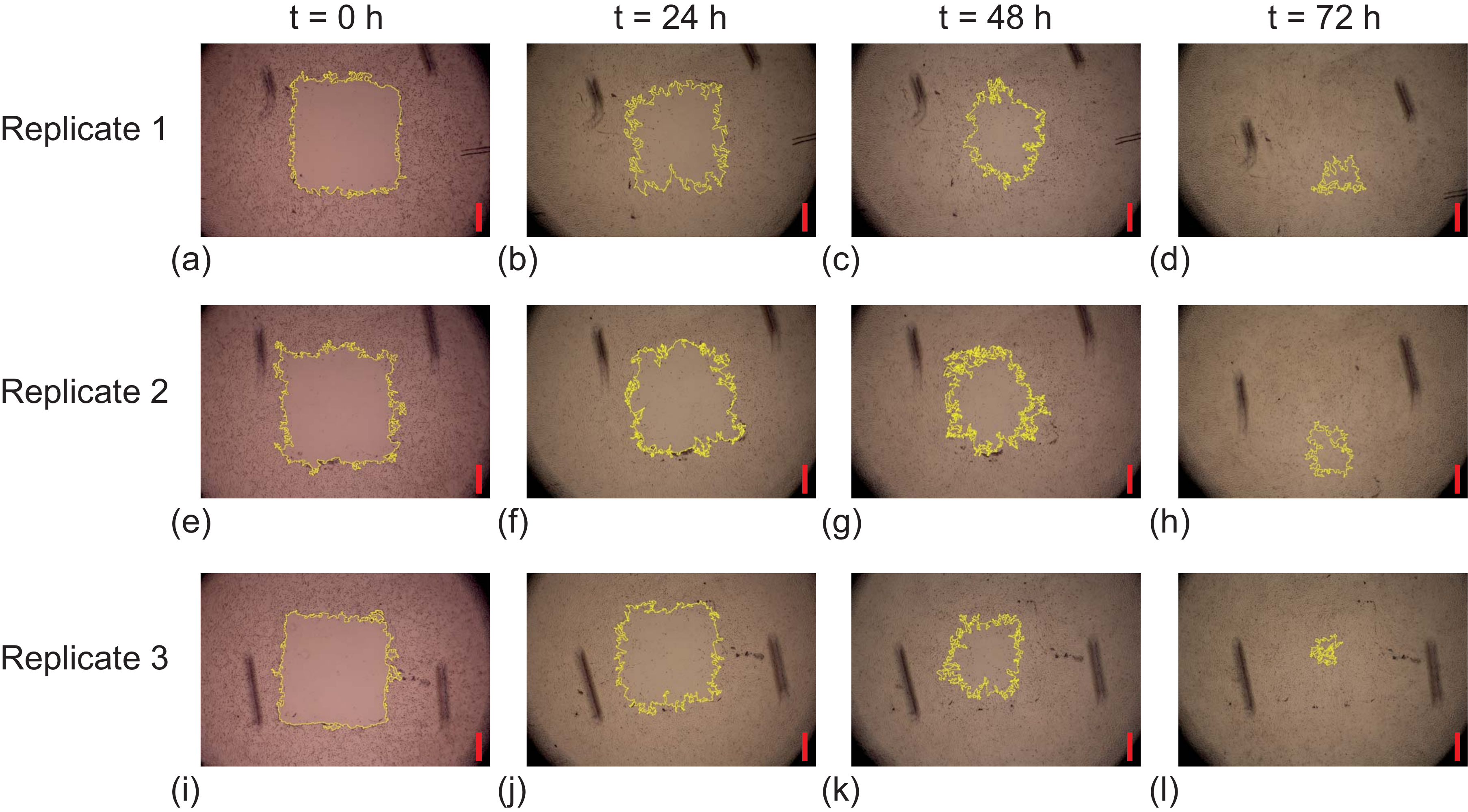}
\caption{{\bf Experimental images of square wound shape with detected wound area.}
(a)--(d) Replicate 1. (e)--(h) Replicate 2. (i)--(l) Replicate 3. The solid yellow line represents the detected wound edge using ImageJ (Ferreira and Rasband, 2012). The scale bar corresponds to 500 $\mu$m}
\label{SF1}
\end{center}
\end{figure}
\end{landscape}

\begin{landscape}
\begin{figure}
\begin{center}
\includegraphics[width=1.6\textwidth]{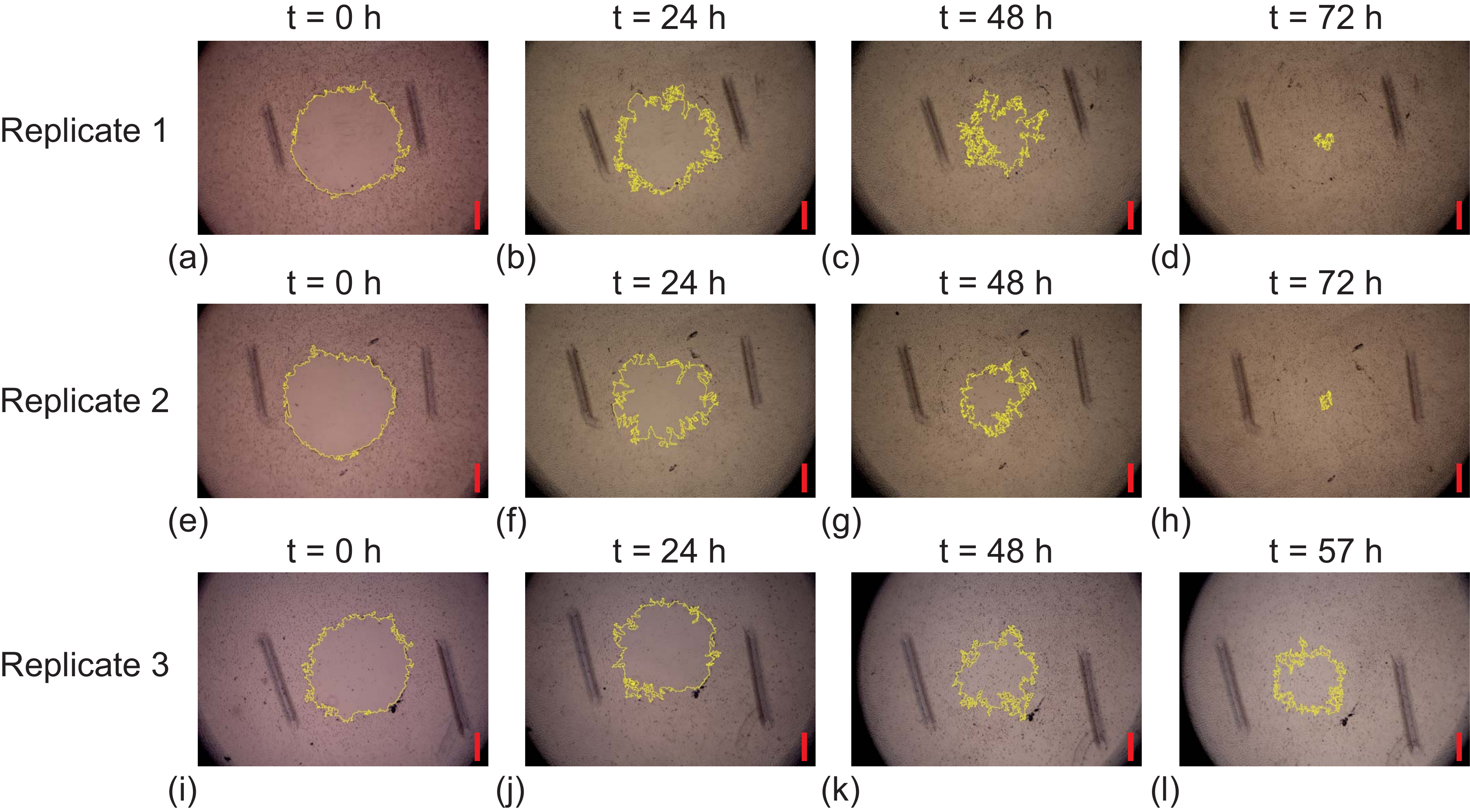}
\caption{{\bf Experimental images of circular wound shape with detected wound area.}
(a)--(d) Replicate 1. (e)--(h) Replicate 2. (i)--(l) Replicate 3. The solid yellow line represents the detected wound edge using ImageJ (Ferreira and Rasband, 2012). The scale bar corresponds to 500 $\mu$m}
\label{SF2}
\end{center}
\end{figure}
\end{landscape}

\begin{landscape}
\begin{figure}
\begin{center}
\includegraphics[width=1.6\textwidth]{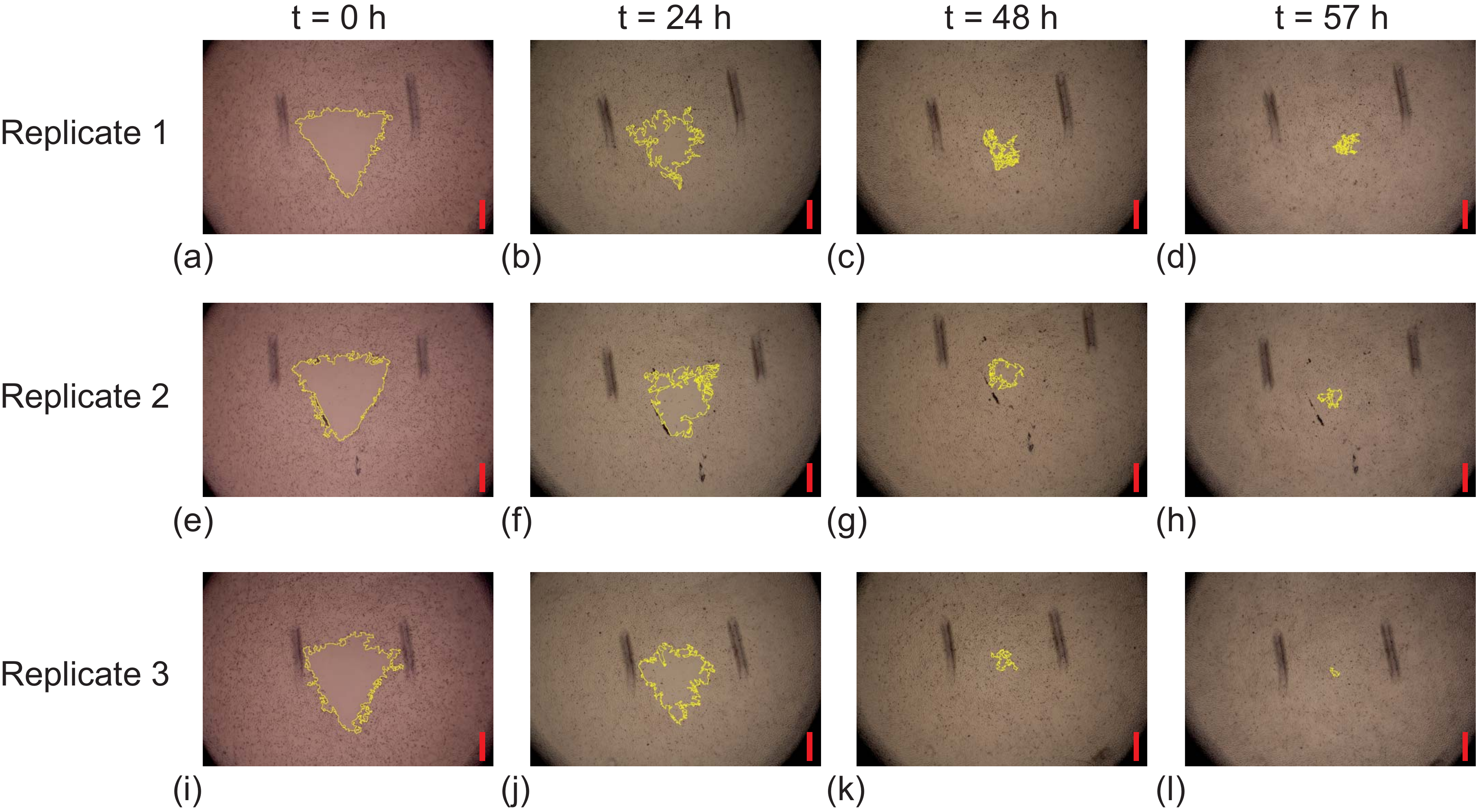}
\caption{{\bf Experimental images of triangular wound shape with detected wound area.}
(a)--(d) Replicate 1. (e)--(h) Replicate 2. (i)--(l) Replicate 3. The solid yellow line represents the detected wound edge using ImageJ (Ferreira and Rasband, 2012). The scale bar corresponds to 500 $\mu$m}
\label{SF3}
\end{center}
\end{figure}
\end{landscape}

We show the experimental images with the detected wound edge at $t = 0, 24, 48, 72$ h in Figs. \ref{SF1}--\ref{SF3}. The estimated wound areas at all the recorded experimental time points, i.e. $t = 0, 9, 24, 33, 48, 57, 72$ h, are listed in Tables \ref{TableAreaS}--\ref{TableAreaT} for the square, circular, and triangular wound shapes, respectively.
\newpage

\newpage

\newpage

\newpage
\begin{table}[ht]
\caption{{\bf Estimates of wound area during the square wound closing.} All estimates are given to three decimal places.}
\renewcommand{\arraystretch}{1.25}
\centering
\begin{tabular}{ | c | c | c | c | c |}
\hline
\multirow{2}{*}{Time (h)} & \multicolumn{4}{|c|}{Wound area (mm$^2$)} \\ [0.5ex]
\cline{2-5}
 & Replicate 1 &  Replicate 2 & Replicate 3 & Average \\
\hline
0 & 4.422 & 4.404 & 4.253 & 4.360\\
\hline
9 & 3.846 & 3.899 & 3.959 & 3.901\\
\hline
24 & 2.945 & 3.225 & 3.182 & 3.117\\
\hline
33 & 2.879 & 3.040 & 2.767 & 2.895\\
\hline
48 & 2.287 & 2.313 & 1.543 & 2.048\\
\hline
57 & 1.24 & 1.17 & 0.991 & 1.134\\
\hline
72 & 0.292 & 0.332 & 0.206 & 0.277\\
\hline
\end{tabular}
\label{TableAreaS}
\end{table}

\begin{table}[ht]
\caption{{\bf Estimates of wound area during the circular wound closing.} All estimates are given to three decimal places. Data for replicate 3 ends at 57 h, and no data is presented at 72 h.}
\renewcommand{\arraystretch}{1.25}
\centering
\begin{tabular}{ | c | c | c | c | c |}
\hline
\multirow{2}{*}{Time (h)} & \multicolumn{4}{|c|}{Wound area (mm$^2$)} \\ [0.5ex]
\cline{2-5}
 & Replicate 1 &  Replicate 2 & Replicate 3 & Average \\
\hline
0 & 3.305 & 3.174 & 2.916 & 3.132\\
\hline
9 & 3.012 & 3.009 & 2.849 & 2.957\\
\hline
24 & 2.497 & 2.105 & 2.555 & 2.386\\
\hline
33 & 2.052 & 1.855 & 2.101 & 2.003\\
\hline
48 & 1.093 & 1.040 & 1.449 & 1.194\\
\hline
57 & 0.666 & 0.525 & 1.099 & 0.763\\
\hline
72 & 0.034 & 0.027 & -  & 0.031\\
\hline
\end{tabular}
\label{TableAreaC}
\end{table}

\newpage
\begin{table}[ht]
\caption{{\bf Estimates of wound area during the triangular wound closing.} All estimates are given to three decimal places.}
\renewcommand{\arraystretch}{1.25}
\centering
\begin{tabular}{ | c | c | c | c | c |}
\hline
\multirow{2}{*}{Time (h)} & \multicolumn{4}{|c|}{Wound area (mm$^2$)} \\ [0.5ex]
\cline{2-5}
 & Replicate 1 &  Replicate 2 & Replicate 3 & Average \\
\hline
0 & 1.569 & 1.756 & 1.743 & 1.689\\
\hline
9 & 1.363 & 1.487 & 1.411 & 1.420\\
\hline
24 & 0.851 & 0.963 & 1.060 & 0.958\\
\hline
33 & 0.561 & 0.786 & 0.773 & 0.707\\
\hline
48 & 0.164 & 0.193 & 0.063 & 0.140\\
\hline
57 & 0.055 & 0.068 & 0.010 & 0.044\\
\hline
\end{tabular}
\label{TableAreaT}
\end{table}

\section{Experimental data: Cell density information}
In Table \ref{TableCellDensityInfo} we show the cell density information in the three chosen subregions in the proliferation assay, which is shown in Fig. 1 in the main manuscript. Each subregion has dimensions 900 $\mu$m $\times$ 300 $\mu$m. Cells in each subregion are counted in Photoshop using the `Count Tool' (Adobe Systems Incorporated, 2017). After counting the number of cells in each subregion, we divide the total number of cells by the total area to estimate the cell density at $t$ = 0, 9, 24, 33, 48, 57, 72, 81 and 96 h.
\begin{landscape}
\begin{table}[ht]
\caption{{\bf Cell density information in the three subregions in the proliferation assay.} All density estimates are given to two significant figures.}
\renewcommand{\arraystretch}{1.25}
\centering
\begin{tabular}{ | c | c | c | c | c | c | c | c | c | c | c |}
\hline
\multicolumn{2}{|c|}{Time (h)} & 0  & 9  & 24  & 33  & 48  & 57  & 72  & 81  & 96 \\ [0.5ex]
\hline
\multirow{2}{*}{Yellow box} & Number of cells & 195 & 202 & 298 & 291 & 321 & 288 & 342 & 317 & 404\\
\cline{2-11}
& Cell density ($\times$ 10$^{-3}$ cells/$\mu$m$^2$) & 0.72 & 0.75 & 1.10 & 1.08 & 1.19 & 1.07 & 1.27 & 1.17 & 1.50\\
\hline
\multirow{2}{*}{Red box} & Number of cells & 175 & 182 & 231 & 294 & 308 & 378 & 368 & 351 & 290\\
\cline{2-11}
& Cell density ($\times$ 10$^{-3}$ cells/$\mu$m$^2$) & 0.65 & 0.67 & 0.86 & 1.09 & 1.14 & 1.40 & 1.36 & 1.30 & 1.07\\
\hline
\multirow{2}{*}{Blue box} & Number of cells & 192 & 211 & 291 & 286 & 288 & 338 & 403 & 391 & 330\\
\cline{2-11}
& Cell density ($\times$ 10$^{-3}$ cells/$\mu$m$^2$) & 0.71 & 0.78 & 1.08 & 1.06 & 1.07 & 1.25 & 1.49 & 1.45 & 1.22\\
\hline
\multirow{2}{*}{Average} & Number of cells & 185 & 192 & 265 & 293 & 315 & 333 & 355 & 334 & 347\\
\cline{2-11}
& Cell density ($\times$ 10$^{-3}$ cells/$\mu$m$^2$) & 0.69 & 0.71 & 0.98 & 1.08 & 1.16 & 1.23 & 1.31 & 1.24 & 1.29\\
\hline
\end{tabular}
\label{TableCellDensityInfo}
\end{table}
\end{landscape}

\subsection{Estimates of contour level}
\begin{landscape}
\begin{figure}[p]
\centering
\includegraphics[width=1.5\textwidth]{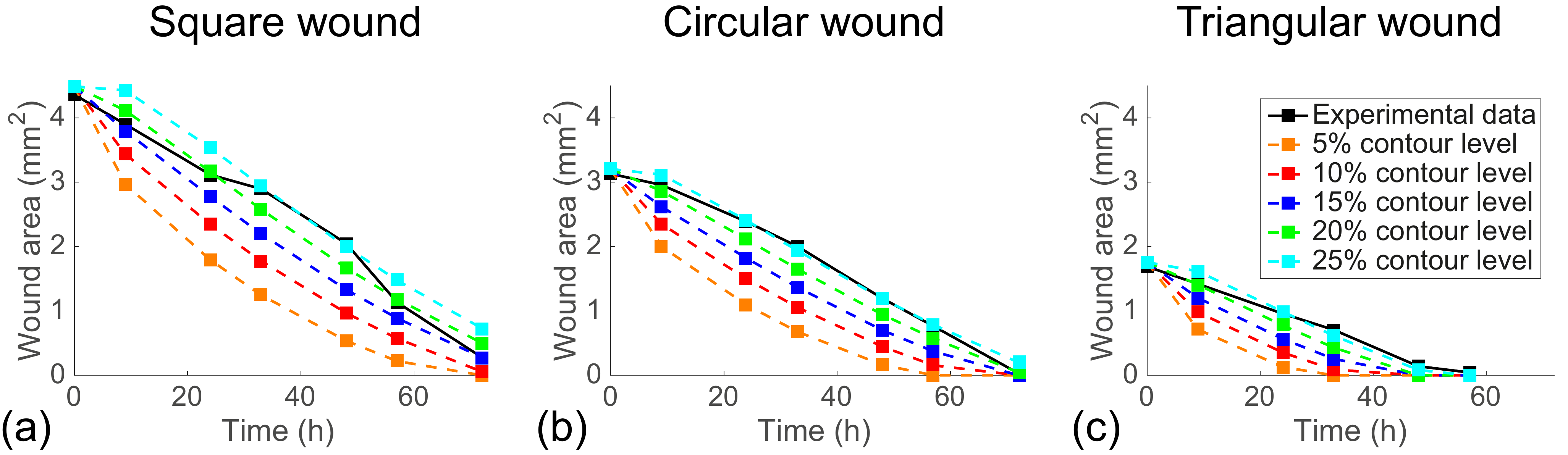}
\caption{{\bf Comparison of the time evolution of the wound area.} The wound area for, (a) square, (b) circular, and (c) triangular wound shape is given. The wound data is given for the averaged experimental wound area and the area estimated from the contours $\mathcal{C}(x,y,t) = 0.05K, 0.1K, 0.15K, 0.2K$, and $0.25K$. }
\label{SF4}
\end{figure}
\end{landscape}

We numerically solve Eq. (\ref{PDE2d-2}) with $D = 1200$ $\mu$m$^2$/h, $r = 0.036$ /h, $K = 1.4 \times 10^{-3}$ cells/$\mu$m$^2$, $\delta  = 10$ $\mu$m, and $\epsilon = 1 \times 10^{-5}$, and measure the area enclosed by various choices of contours using the ImageJ edge detection algorithm (Ferreira and Rasband, 2012). The estimated wound areas from the contours $\mathcal{C}(x,y,t) = 0.05K, 0.1K, 0.15K, 0.2K$, and $0.25K$ for the three wounds are listed in Tables \ref{TableContourS}--\ref{TableContourT}, respectively. The time evolution of the area enclosed by the contours are superimposed on the time evolution of the averaged experimental wound area, shown in Figure \ref{SF4}.

We then measure the least--squares error between the area enclosed by the contour of Eq. (\ref{PDE2d-2}) and averaged experimental wound area, given by
\begin{align}
E(\mathcal{C}(x,y,t)) = & \sum_{l=1}^{L}\left[ A^{\textrm{model}}(\mathcal{C}(x,y,t_l)) - A^{\textrm{data}}(\mathcal{C}(x,y,t_l)) \right]_{\text{Square}}^2 + \nonumber\\ & \sum_{l=1}^{L}\left[ A^{\textrm{model}}(\mathcal{C}(x,y,t_l)) - A^{\textrm{data}}(\mathcal{C}(x,y,t_l)) \right]_{\text{Circle}}^2 + \nonumber\\ & \sum_{l=1}^{L}\left[ A^{\textrm{model}}(\mathcal{C}(x,y,t_l)) - A^{\textrm{data}}(\mathcal{C}(x,y,t_l)) \right]_{\text{Triangle}}^2,
\end{align}
where $\mathcal{C}(x,y,t)$ is the contour level, $A^{\textrm{model}}(\mathcal{C}(x,y,t_l))$ is the wound area estimated from the contour of Eq. (\ref{PDE2d-2}) at time $t_l$, and $A^{\textrm{data}}(\mathcal{C}(x,y,t_l))$ is the averaged wound area estimated from the experimental image at time $t_l$.  Here,  $l = 1,2, \ldots, L$ is an index indicating the number of time points used to compare the experimental wound area with the area enclosed by the contours. We then identify the contour $\mathcal{C}^*(x,y,t) = 0.2K$, which minimises the least--squares measure, $E_{\text{min}}=E(C^*)$, for all the three wounds.
\newpage
\begin{landscape}
\begin{table}[ht]
\caption{{\bf Estimates of wound area using various contour levels during the square wound closing.} All estimates are given to three decimal places.}
\renewcommand{\arraystretch}{1.25}
\centering
\begin{tabular}{ | c | c | c | c | c | c |}
\hline
\multirow{2}{*}{Time (h)} & \multicolumn{5}{|c|}{Wound area (mm$^2$)} \\ [0.5ex]
\cline{2-6}
 & $\mathcal{C}(x,y,t) = 0.05K$ &  $\mathcal{C}(x,y,t) = 0.1K$ & $\mathcal{C}(x,y,t) = 0.15K$ & $\mathcal{C}(x,y,t) = 0.2K$ & $\mathcal{C}(x,y,t) = 0.25K$\\
\hline
0 & 4.489 & 4.489 & 4.489 & 4.489 & 4.489\\
\hline
9 & 2.971 & 3.437 & 3.794 & 4.116 & 4.428\\
\hline
24 & 1.793 & 2.345 & 2.778 & 3.163 & 3.535\\
\hline
33 & 1.246 & 1.775 & 2.198 & 2.578 & 2.942\\
\hline
48 & 0.536 & 0.968 & 1.332 & 1.668 & 1.995\\
\hline
57 & 0.220 & 0.573 & 0.887 & 1.184 & 1.477\\
\hline
72 & 0 & 0.056 & 0.277 & 0.5 & 0.728\\
\hline
\end{tabular}
\label{TableContourS}
\end{table}
\end{landscape}

\begin{landscape}
\begin{table}[ht]
\caption{{\bf Estimates of wound area using various contour levels during the circular wound closing.} All estimates are given to three decimal places.}
\renewcommand{\arraystretch}{1.25}
\centering
\begin{tabular}{ | c | c | c | c | c | c |}
\hline
\multirow{2}{*}{Time (h)} & \multicolumn{5}{|c|}{Wound area (mm$^2$)} \\ [0.5ex]
\cline{2-6}
 & $\mathcal{C}(x,y,t) = 0.05K$ &  $\mathcal{C}(x,y,t) = 0.1K$ & $\mathcal{C}(x,y,t) = 0.15K$ & $\mathcal{C}(x,y,t) = 0.2K$ & $\mathcal{C}(x,y,t) = 0.25K$\\
\hline
0 & 3.205 & 3.205 & 3.205 & 3.205 & 3.205\\
\hline
9 & 2.001 & 2.349 & 2.618 & 2.862 & 3.118\\
\hline
24 & 1.091 & 1.496 & 1.819 & 2.111 & 2.406\\
\hline
33 & 0.677 & 1.055 & 1.366 & 1.650 & 1.936\\
\hline
48 & 0.161 & 0.445 & 0.700 & 0.943 & 1.189\\
\hline
57 & 0 & 0.155 & 0.372 & 0.572 & 0.786\\
\hline
72 & 0 & 0 & 0 & 0.045 & 0.200\\
\hline
\end{tabular}
\label{TableContourC}
\end{table}
\end{landscape}

\begin{landscape}
\begin{table}[ht]
\caption{{\bf Estimates of wound area using various contour levels during the triangular wound closing.} All estimates are given to three decimal places.}
\renewcommand{\arraystretch}{1.25}
\centering
\begin{tabular}{ | c | c | c | c | c | c |}
\hline
\multirow{2}{*}{Time (h)} & \multicolumn{5}{|c|}{Wound area (mm$^2$)} \\ [0.5ex]
\cline{2-6}
 & $\mathcal{C}(x,y,t) = 0.05K$ &  $\mathcal{C}(x,y,t) = 0.1K$ & $\mathcal{C}(x,y,t) = 0.15K$ & $\mathcal{C}(x,y,t) = 0.2K$ & $\mathcal{C}(x,y,t) = 0.25K$\\
\hline
0 & 1.757 & 1.757 & 1.757 & 1.757 & 1.757\\
\hline
9 & 0.713 & 0.981 & 1.203 & 1.408 & 1.616\\
\hline
24 & 0.129 & 0.354 & 0.566 & 0.775 & 0.989\\
\hline
33 & 0 & 0.088 & 0.255 & 0.431 & 0.617\\
\hline
48 & 0 & 0 & 0 & 0 & 0.082\\
\hline
57 & 0 & 0 & 0 & 0 & 0\\
\hline
\end{tabular}
\label{TableContourT}
\end{table}
\end{landscape}

